\documentclass[prc,twocolumn,superscriptaddress,amssymb,amsmath,amsfonts,aps]{revtex4}
\usepackage{graphicx}
\usepackage{dcolumn}

\begin{document}
\title{\large On the definitions of
the $\gamma^* N\rightarrow N^*$ helicity amplitudes}

\newcommand*{\JLAB}{Thomas Jefferson National Accelerator Facility,
Newport News, Virginia 23606}
\affiliation{\JLAB}
\newcommand*{\YEREVAN}{Yerevan Physics Institute, 375036 Yerevan,
Armenia}
\affiliation{\YEREVAN}
\newcommand*{\ANL}{Argonne National Laboratory, 
Argonne,
Illinois 60439}
\affiliation{\ANL}
\author{I.G.~Aznauryan}
     \affiliation{\JLAB}
     \affiliation{\YEREVAN}
\author{V.D.~Burkert}
     \affiliation{\JLAB}
\author{T.-S.H.~Lee}
     \affiliation{\JLAB}
     \affiliation{\ANL}

\begin{abstract}
{We present definitions and formulas
that may be useful for a consistent computation of 
helicity amplitudes
for the process $\gamma^* N\rightarrow N^*$ 
in theoretical approaches. Of particular importance is the 
correct determination
of the common sign of the amplitudes and of the relative
sign between the transverse ($A_{1/2},A_{3/2}$) and longitudinal
($S_{1/2}$) amplitudes. This clarification is necessary to clear up  
confusions present in theoretical works. Using the definitions
presented in this paper will enable
a direct comparison with 
amplitudes extracted from experimental data.
}
\end{abstract}
\maketitle
\section{Introduction}
The excitation of nucleon resonances
in electromagnetic interactions has long been known
as an important source of information on the baryon 
structure and long- and short-range interaction 
in the domain of quark confinement.
Constituent quark models have been developed
that relate electromagnetic resonance transition form factors
to fundamental quantities, such as the quark confining potential.
While the picture of the $N$ and $N^*$ built from 3 quarks 
is recognized as a basic starting point in the description
of these states, the inter-quark interaction does not
exclude the possibility of additional degrees
of freedom, namely, the presence of $3q-q\bar q$ components
in the $N$ and $N^*$. There is also a possibility of
alternative structures, such as hybrid $q^3G$ states and
resonances dynamically generated in the meson-baryon interaction.

The $Q^2$ dependence of the  $\gamma^* N\rightarrow N^*$ transition
amplitudes is highly sensitive to different
descriptions of the nucleon resonances. Rich
information on these amplitudes, 
both transverse and longitudinal, has become 
available in a wide $Q^2$ range
for the N(1440)P$_{11}$,
N(1520)D$_{13}$, N(1535)S$_{11}$ resonances 
due to the recent high precision JLab-CLAS measurements 
of the $\vec{e}p\rightarrow ep\pi^0,en\pi^+$ reactions
\cite{Joo1,Joo2,Joo3,Egiyan,Park,Azn04,Azn065,AznRoper}. 
More results
are expected in $\pi$ and $2\pi$ electroproduction.        

While comparing the predictions obtained in different
approaches with the $\gamma^* N\rightarrow N^*$ amplitudes
extracted from experimental data, we found that there are
two points which introduce confusion in theoretical
predictions and in most cases
do not allow to compare properly the results of
different approaches with each other and with experimental
data:

(i) It is known that experimental results on
the $\gamma^* N\rightarrow N^*$ helicity amplitudes
$A_{1/2},~A_{3/2},~S_{1/2}$,
extracted from the contribution of the diagram
of Fig. 1 to $\gamma^* N\rightarrow N\pi$, contain
the sign of the $\pi N N^*$ vertex. However, by the definition
accepted many years ago,
this fact is not reflected explicitly in the amplitudes
extracted from experiment. In many cases 
this is not taken into account in theoretical calculations and
causes difficulties in dealing with the common
sign of the predicted amplitudes.

(ii) The definition of the amplitudes $A_{1/2},~A_{3/2},~S_{1/2}$
through the hadron electromagnetic current 
which is commonly used in theoretical approaches
contains points
which can introduce mistakes in the relative sign of
longitudinal amplitude $S_{1/2}$ relative to the 
transverse ones $A_{1/2},~A_{3/2}$.
This is the second source of difficulties in dealing
with theoretical predictions.
 
\begin{figure}[ht]
\includegraphics[width=17.cm, bb=-45 640 596 720]{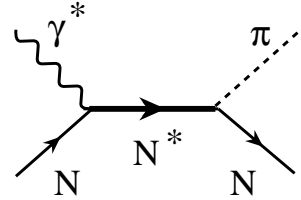}
\caption{The diagram corresponding to the
$N^*$ contribution to $\gamma^* N\rightarrow N\pi$.
}
\label{fig:fig1}
\end{figure}

Our goal in this paper is to present different definitions
of the $\gamma^* p\rightarrow N^{*+}$
helicity amplitudes: through the $\gamma^* p\rightarrow N\pi$
multipole amplitudes, through the hadron electromagnetic current,
in terms of the $\gamma^* p\rightarrow N^{*+}$
form factors, and in quark model,
which are consistent with each other.
To exclude possible sources of mistakes,
we give explicitly the definitions of all quantities
which enter the formulas.

We pay special attention to the way, how the sign
of the $\pi N N^*$ vertex should be taken into account.
It is known that this sign comes from the
relative contributions of the diagrams which present
the resonance (Fig. 1)
and Born terms (Fig. 2) contributions 
to $\gamma^* N\rightarrow N\pi$.
We will give formulas which 
relate the $\gamma^* p\rightarrow N^{*+}$ helicity
amplitudes extracted from experiment to those 
which are calculated
through the hadron electromagnetic current and are multiplied by
the sign of the ratio of the coupling constants
in the vertices $\pi N N$ and $\pi N N^*$.

\begin{figure}[ht]
\includegraphics[width=11.8cm, bb=35 580 596 800]{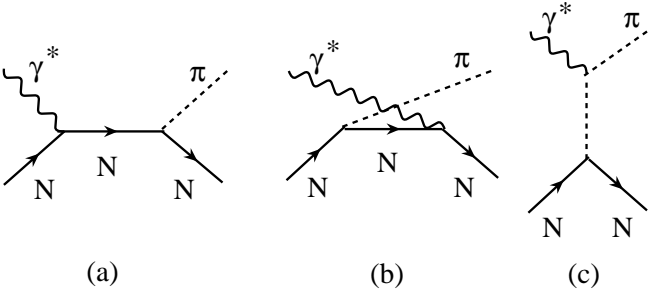}
\caption{The diagrams corresponding to the Born terms
contributions to $\gamma^* N\rightarrow N\pi$.
}
\label{fig:fig2}
\end{figure}

In Sec. II, we present the definitions of the 
$\gamma^* p\rightarrow N^{*+}$
helicity amplitudes
through the $\gamma^* p\rightarrow N\pi$
multipole amplitudes, which are used in the extraction
of $A_{1/2},~A_{3/2},~S_{1/2}$ from the
$\gamma^* p\rightarrow N\pi$ data.
These definitions are well known and coincide, for example,
with the definitions presented in Refs. 
\cite{PDG,Arndt,Capstick,Kamalov}.

In Sec. III, we present the definition of the 
$\gamma^* p\rightarrow N^{*+}$
helicity amplitudes
through the hadron electromagnetic current.
To distinguish between these amplitudes and those extracted
from experiment, we denote the amplitudes
defined through the hadron electromagnetic current by
 ${\cal A}_{1/2},~{\cal A}_{3/2},~{\cal S}_{1/2}$: 
\begin{equation}
A_{\frac{1}{2},\frac{3}{2}} 
= \zeta {\cal A}_{\frac{1}{2},\frac{3}{2}},
~~~S_{\frac{1}{2}}=\zeta {\cal S}_{\frac{1}{2}}.
\end{equation}
Here $\zeta$ is the sign which reflects the presence
of the $\pi NN^*$ vertex in Fig. 1.

In Sec. IV, using the definitions of 
Sec. III, we present the expressions for the 
amplitudes ${\cal A}_{1/2},~{\cal A}_{3/2},~{\cal S}_{1/2}$ 
in terms of the $\gamma^* N\rightarrow N^{*}$ form factors.

In Sec. V, the relation between $\zeta$ and the
sign of the ratio of the 
$\pi NN$, $\pi NN^*$ coupling constants
is found from the covariant
calculations of the Figs. 1,2 contributions to
$\gamma^* N\rightarrow N\pi$.

In Sec. VI, we present formulas for the calculation
of the 
amplitudes ${\cal A}_{1/2},~{\cal A}_{3/2},~{\cal S}_{1/2}$
in nonrelativistic quark model. To avoid possible
sources of mistakes, we give explicitly the definitions
of all quantities which enter these formulas and present
in explicit form definitions of wave functions
for the $N$, $P_{11}(1440)$, and $S_{11}(1535)$.
Further, in Sec. VI, we present formulas for the
amplitudes ${\cal A}_{1/2},~{\cal S}_{1/2}$ for the
transitions  $\gamma^* p\rightarrow P_{11}(1440),S_{11}(1535)$
in nonrelativistic quark model.

In Sec. VII, we present information on the signs of the $\pi N N^*$ 
coupling constants available in quark model,
and demonstrate the importance of using these
signs for the correct presentation of quark model predictions.
In other approaches, such as
dynamically generated resonances, these signs
should be found within these approaches.
Only in this case, we can make proper comparison
of the predictions obtained in different approaches
with each other and with the results extracted from
experiment. 

\section{Definition of the $\gamma^* p\rightarrow N^{*+}$ 
helicity amplitudes 
through the $\gamma^* p\rightarrow N\pi$ 
multipole amplitudes}

The $\gamma^* N\rightarrow N^{*}$
helicity amplitudes, extracted from experimental data
on the reaction $eN\rightarrow eN\pi$, are usually presented
through the $\gamma^* N\rightarrow N\pi$
multipole amplitudes. In the case of 
the $\gamma^* p\rightarrow N^{*+}$
amplitudes, the corresponding formulas are following.

For $l+$ multipole amplitudes:
\begin{eqnarray}
&&A_{1/2}=-\frac{1}{2}\left[(l+2){\cal E}_{l+}
+l{\cal M}_{l+}\right], \\
&&A_{3/2}=\frac{\left[l(l+2)\right]^{1/2}}{2}
({\cal E}_{l+}-{\cal M}_{l+}),\\
&&S_{1/2}=-\frac{1}{\sqrt{2}}(l+1){\cal S}_{l+}.
\end{eqnarray}
For $(l+1)-$ multipole amplitudes:
\begin{eqnarray}
&&A_{1/2}=\frac{1}{2}\left[
(l+2){\cal M}_{(l+1)-}-l{\cal E}_{(l+1)-}\right],\\
&&A_{3/2}=-\frac{\left[l(l+2)\right]^{1/2}}{2}
({\cal E}_{(l+1)-}+{\cal M}_{(l+1)-}),\\
&&S_{1/2}=-\frac{1}{\sqrt{2}}(l+1)
{\cal S}_{(l+1)-}.
\end{eqnarray}
Here ${\cal M}_{l\pm}({\cal E}_{l\pm},{\cal S}_{l\pm})$
are the amplitudes which are related to the resonance
contributions to the multipole amplitudes at the resonance
positions via:

\begin{eqnarray}
&&
ImM_{l\pm}(E_{l\pm},S_{l\pm})(W=M)
\equiv
a{\cal M}_{l\pm}({\cal E}_{l\pm},{\cal S}_{l\pm}),
\\
&&a\equiv C_I
\left[\frac{1}{(2J+1)\pi}\frac{K}{q_r}\frac{m}{M}
\frac{\beta_{\pi N}}{\Gamma}\right]^{1/2},\\
&&C_{1/2}=-\sqrt{\frac{1}{3}},~C_{3/2}=\sqrt{\frac{2}{3}}
~for~\gamma^* p\rightarrow \pi^0 p,\\
&&C_{1/2}=-\sqrt{\frac{2}{3}},~C_{3/2}=-\sqrt{\frac{1}{3}}
~for~\gamma^* p\rightarrow \pi^+ n.
\end{eqnarray}
In Eq. (9), $\Gamma$, $M$, $J$ and $I$ are the total width, mass,
spin and isospin of the
resonance, $\beta_{\pi N}$ is its branching ratio
to the $\pi N$ channel, $m$ is the nucleon mass,
$K\equiv (M^2-m^2)/2M$ and $q_r$ are momenta
of the real photon and pion at the resonance position 
in the c.m.s.

With this definition, the $N^*\rightarrow N\gamma$ width
is
\begin{equation}
\Gamma(N^*\rightarrow N\gamma)=
\frac{2K^2}{\pi(2J+1)}\frac{m}{M}\left(|A_{1/2}|^2+|A_{3/2}|^2\right).
\end{equation}

Below we give the relations between
multipole
amplitudes $M_{l\pm}(W,Q^2)$, $E_{l\pm}(W,Q^2)$,
$S_{l\pm}(W,Q^2)$ and the $\gamma^*N\rightarrow \pi N$
differential cross section. It is convenient to
present these relations in terms of the intermediate  amplitudes
$F_{1,2,...6}(W,cos \theta,Q^2)$ :
\begin{eqnarray}
F_1=&&\sum\{(lM_{l+}+E_{l+})P'_{l+1}(x)\nonumber\\
&&+[(l+1)M_{l-}+E_{l-}]P'_{l-1}(x)\},\\
F_2=&&\sum[(l+1)M_{l+}+lM_{l-}] P'_l(x), \\
F_3=&&\sum[(E_{l+}-M_{l+})P''_{l+1}(x)\nonumber\\
&&+(E_{l-}+M_{l-})P''_{l-1}(x)],\\
F_4=&&\sum(M_{l+}-E_{l+}-M_{l-}-E_{l-})P''_l(x), \\
F_5=&&\sum[(l+1)S_{l+}P'_{l+1}(x)-
lS_{l-}P'_{l-1}(x)],\\
F_6=&&\sum[lS_{l-}-(l+1)S_{l+}] P'_l(x).
\end{eqnarray}
The amplitudes
$F_{1,2,...6}(W,cos \theta,Q^2)$ are related to the helicity
amplitudes and the cross section 
of the $\gamma^*N\rightarrow \pi N$ reaction in the following way:
\begin{eqnarray}
&&H_1=\frac{-1}{\sqrt{2}}sin\theta(F_3+F_4cos\theta),\\
&&H_2=\frac{-1}{\sqrt{2}}sin\theta(2F_1-2F_2cos\theta
+F_4sin^2\theta),\\
&&H_3=\frac{-1}{\sqrt{2}}F_4sin^2\theta,\\
&&H_4=\frac{1}{\sqrt{2}}sin\theta(2F_2+F_3+F_4cos\theta),\\
&&H_5=\frac{Q}{|\bf{k}|}(F_5+F_6cos\theta),\\
&&H_6=\frac{Q}{|\bf{k}|}F_6sin\theta,
\end{eqnarray}
\begin{eqnarray}
\frac{d\sigma}{d\Omega_{\pi}}=&&\sigma_T+\epsilon\sigma_L
+\epsilon\sigma_{TT}cos2\phi\nonumber\\
&&+\sqrt{2\epsilon(1+\epsilon)}\sigma_{LT}cos\phi,
\end{eqnarray}
\begin{eqnarray}
&\sigma_T=\frac{|\bf{q}|}{2K}(|H_1|^2+|H_2|^2+|H_3|^2+|H_4|^2),\\
&\sigma_L=\frac{|\bf{q}|}{K}(|H_5|^2+|H_6|^2),\\
&\sigma_{TT}=\frac{|\bf{q}|}{K}Re(H_3H_2^*-H_4H_1^*),\\
&\sigma_{LT}=\frac{|\bf{q}|}{\sqrt{2}K}
Re[(H_1-H_4)H_5^*+(H_2+H_3)H_6^*].
\end{eqnarray}

In Eqs.(13-29), 
$\bf{k}$ and $\bf{q}$ are the momenta of the virtual photon
and pion in the c.m.s. of the reaction $\gamma^*N\rightarrow \pi N$, 
$\theta$ and $\phi$ are the polar and 
azimuthal angles of the pion,
$\epsilon$ is the polarization
factor of the virtual photon, 
$x\equiv cos\theta$.

\section{Definition of the $\gamma^* N\rightarrow N^{*}$ 
helicity amplitudes 
through the hadron electromagnetic current}

The definition of the 
$\gamma N \rightarrow N^*$ helicity amplitudes
through the hadron electromagnetic current,
which is commonly used for the calculation
of these amplitudes 
in theoretical approaches, is:

\begin{eqnarray}
&&{\cal A}_{\frac{1}{2}}=\sqrt{\frac{2\pi \alpha}{K}}\frac{1}{e}
<S_z^*=\frac{1}{2}|\epsilon^{(+)}_{\mu}J^{\mu}|S_z=-\frac{1}{2}>,\\
&&{\cal A}_{\frac{3}{2}}=\sqrt{\frac{2\pi \alpha}{K}}\frac{1}{e}
<S_z^*=\frac{3}{2}|\epsilon^{(+)}_{\mu}J^{\mu}|S_z=\frac{1}{2}>,\\
&&{\cal S}_{\frac{1}{2}}=\sqrt{\frac{2\pi \alpha}{K}}
\frac{1}{e}
<S_z^*=\frac{1}{2}|\frac{|{\bf{k}}|}{Q}
\epsilon^{(0)}_{\mu}J^{\mu}|S_z=\frac{1}{2}>.
\end{eqnarray}
In order to avoid the mistakes in using these
formulas, below 
we give explicitly the definitions of all quantities
which enter Eqs. (30-32).

Let us denote the 4-momenta of the virtual photon, nucleon and 
resonance in the vertex $\gamma N 
\rightarrow N^*$ through $k,p,p^*$, respectively:
\begin{equation}
p^*=p+k.
\end{equation}

The z-axis is directed along the photon 3-momentum
(${\bf{k}}$) in the $N^*$ rest frame,  $Q\equiv\sqrt{-k^2}$,
and $S_z,S^*_z$ are the projections of the nucleon
and resonance spins on the z-axis.

With the definition $a_{\mu}\equiv (a_0,-{\bf{a}})$, we have
\begin{equation}
\epsilon^{(0)}_{\mu}=\frac{1}{Q}(|{\bf{k}}|,0,0,-k_0),
\end{equation}

$~~~~~~~~~~~\epsilon^{(+)}_{\mu}$=(0,~-~{\boldmath$\epsilon$}$^{(+)}$),
{\boldmath$\epsilon$}$^{(+)}=-\frac{1}{\sqrt{2}}(1,i,0),~~~~~~~$ (34')

and
\begin{eqnarray}
&&\epsilon^{(+)}_{\mu}J^{\mu}=\frac{{\bf{J}}_x+i{\bf{J}}_y}{\sqrt{2}},\\
&&\frac{|{\bf{k}}|}{Q}\epsilon^{(0)}_{\mu}J^{\mu}=J_0,
\end{eqnarray}
where in the last expression we have taken into account
the gauge invariance condition: 
${\bf{J}}_z=J_0\frac{Q}{|{\bf{k}}|}$. 

\section{The $\gamma^* N\rightarrow N^{*}$ 
helicity amplitudes in terms of 
the $\gamma^* N\rightarrow N^{*}$ form factors}

In some theoretical approaches, it is convenient to use
the definition of the $\gamma^* N\rightarrow N^{*}$
helicity amplitudes in terms of
the $\gamma^* N\rightarrow N^{*}$ form factors.
For the $J^P={\frac{1}{2}}^{\pm}$ 
resonances, we will present this definition 
explicitly using the form factors 
introduced in Ref. \cite{Devenish}.

For the $J^P={\frac{1}{2}}^{+}$ 
resonances, the definition \cite{Devenish} is:
\begin{eqnarray}
&&<N^{*}|J_{\mu}|N>\equiv e\bar{u}(p^*)\tilde{J}_{\mu}u(p),\\
&&{\tilde{J}}_{\mu} =
-\left[k^2\gamma_{\mu}-(k\gamma)k_{\mu}\right]G_1(Q^2)\\
&&~~~~~~-\left[(Pk)\gamma_{\mu}-(k\gamma)P_{\mu}\right]G_2(Q^2),\nonumber
\end{eqnarray}

For the $J^P={\frac{1}{2}}^{-}$ 
resonances:
\begin{eqnarray}
&&<N^{*}|J_{\mu}|N>\equiv e\bar{u}(p^*)\tilde{J}_{\mu}\gamma_5 u(p),\\
&&{\tilde{J}}_{\mu} =
\left[k^2\gamma_{\mu}-(k\gamma)k_{\mu}\right]G_1(Q^2)\\
&&~~~~~~+\left[(Pk)\gamma_{\mu}-(k\gamma)P_{\mu}\right]G_2(Q^2),\nonumber
\end{eqnarray}
where $P\equiv \frac{1}{2}(p^*+p)$, and
the $\gamma$ matrices are defined in the following way:
\begin{eqnarray}
&&{\vec\gamma} =
\left(\begin{array}{cc}0&\vec{\sigma}\\-\vec{\sigma}&0\end{array}\right),
~~\gamma_0 =
\left(\begin{array}{cc}1&0\\0&-1\end{array}\right),\\
&&\gamma_5 =
-\left(\begin{array}{cc}0&1\\1&0\end{array}\right),\\
&&\gamma^i=-\gamma_i={\vec\gamma}_i,
~~\gamma_5=i\gamma^0\gamma^1\gamma^2\gamma^3.
\end{eqnarray}

The Dirac equation and the Dirac spinor are:

\begin{eqnarray}
&&(\gamma_{\mu}p^{\mu}-m)u(p)=0, \\
&&u_{s_z}(p)=\sqrt{\frac{E+m}{2m}}\left(\begin{array}{c}
1\\\frac{\bf{\sigma}\bf{p}}{E+m}\end{array}\right)\varphi_{s_z},
\end{eqnarray}
E is the nucleon energy. 

Using the definitions (30,32,37-40), we obtain
the following relations between the $\gamma^* N\rightarrow N^{*}$
helicity amplitudes and form factors.

The $J^P={\frac{1}{2}}^+$ resonances:
\begin{eqnarray}
&&{\cal A}_{\frac{1}{2}}=
\left[2Q^2G_1(Q^2)-(M^2-m^2)G_2(Q^2)\right]b,\\
&&{\cal S}_{\frac{1}{2}}=\tilde {S}b\frac{|{\bf{k}}|}{\sqrt{2}},\\
&&\tilde {S}=2(M+m)G_1(Q^2)+(M-m)G_2(Q^2),\nonumber\\
&&b=e\sqrt{\frac{E-m}{8mK}}.\nonumber
\end{eqnarray}

The $J^P={\frac{1}{2}}^-$ resonances:
\begin{eqnarray}
&&{\cal A}_{\frac{1}{2}}=
\left[2Q^2G_1(Q^2)-(M^2-m^2)G_2(Q^2)\right]b,\\
&&{\cal S}_{\frac{1}{2}}=-\tilde {S}b\frac{|{\bf{k}}|}{\sqrt{2}},\\
&&\tilde {S}=2(M-m)G_1(Q^2)+(M+m)G_2(Q^2),\nonumber\\
&&b=e\sqrt{\frac{E+m}{8mK}}.\nonumber
\end{eqnarray}

For the resonances
with $J\geq \frac{3}{2}$, the relations
between the $\gamma^* N\rightarrow N^{*}$
helicity amplitudes and form factors
can be found using the results of Ref. \cite{Devenish}.
In Ref. \cite{Devenish}, the amplitudes $h_1,h_2,h_3$
are introduced, which are proportional, respectively, to 
${\cal S}_{1/2},~{\cal A}_{3/2},~{\cal A}_{1/2}$;
the relations between $h_1,h_2,h_3$ and
the $\gamma^* N\rightarrow N^{*}$ form factors
are presented too. These relations are quite 
cumbersome. For this reason, here we give only the
relations between 
${\cal A}_{1/2},~{\cal A}_{3/2},~{\cal S}_{1/2}$  
and $h_3,h_2,h_1$, which
for the resonances with $J^P=\frac{3}{2}^{\pm},\frac{5}{2}^{\mp}$,...
have the form:
\begin{eqnarray}
&&{\cal A}_{1/2}=h_3X,\\
&&{\cal A}_{3/2}=\pm \sqrt{3}\frac{h_2}{l}X,\\
&&{\cal S}_{1/2}=h_1\frac{|\bf{k}|}{\sqrt{2}M}X,\\
&&X\equiv \sqrt{\pi\alpha\frac{(M\mp m)^2+Q^2}{24MmK}},\nonumber
\end{eqnarray}
where $l=J-\frac{1}{2}$.

\section{The relation between the $\gamma^* N\rightarrow N^*$
amplitudes extracted from $\gamma^* N\rightarrow N\pi$ and
defined through hadron electromagnetic current }

In this Section, our goal is to find the explicit
relation between the $\gamma^* N\rightarrow N^*$  
form factors contributions to the $\gamma^* 
N\rightarrow N\pi$ multipole amplitudes 
and the  $\gamma^* N\rightarrow N^*$ vertex. 
This will allow us to find the connection between $\zeta$ 
in Eqs. (1) and the sign of the ratio of the 
$\pi NN$, $\pi NN^*$ coupling constants.
This will allow us also to check the
consistency of the relative sign
between longitudinal ($S_{1/2}$) and transverse ($A_{1/2},A_{3/2}$)
amplitudes in the definitions through multipole amplitudes
(2-7) and hadron electromagnetic current (30-32).
With this aim, we will present in detail the results of the covariant
calculations of the Figs. 1,2 contributions to
$\gamma^* N\rightarrow N\pi$ for the resonances
with $J^P={\frac{1}{2}}^{\pm}$.
For the resonances with $J\geq \frac{3}{2}$, the relation
between $\zeta$ and the sign of the ratio of the
$\pi NN$, $\pi NN^*$ coupling constants can be found
from the results of Ref. \cite{Devenish}.
We will present it in the end of this Section
along with that for the $J^P={\frac{1}{2}}^{\pm}$
resonances.

We will use the definitions (37-40)
for the $\gamma^* N\rightarrow N^*$ vertices and will
define the $\pi NN^*$ coupling constants
according to Ref. \cite{Devenish} in the following form. 

The $J^P={\frac{1}{2}}^-$ resonances:
\begin{equation}
<N|J_{\pi}(0)|N^{*+}>=C_I
g^*\bar{u}(p')u(p^*),
\end{equation}

\begin{equation}
\Gamma(N^{*+}\rightarrow N\pi)=\frac{g^{*2}}{4\pi}\frac{E'+m}{M}q_{\pi}.
\end{equation}

The $J^P={\frac{1}{2}}^+$ resonances:
\begin{equation}
<N|J_{\pi}(0)|N^{*+}>=-C_I
g^*\bar{u}(p')\gamma_5 u(p^*),
\end{equation}

\begin{equation}
\Gamma(N^{*+}\rightarrow N\pi)=\frac{g^{*2}}{4\pi}\frac{E'-m}{M}q_{\pi}.
\end{equation}
In Eqs. (54,56), $E'$ and $p'$ are the energy
and 4-momentum of the final nucleon in the reaction 
$\gamma^* N\rightarrow N\pi$ in the c.m.s.,  
$q_{\pi}$ is the pion  3-momentum in this system.

The $\pi NN$ coupling constant is defined according to 

\begin{equation}
<N^+|J_{\pi}(0)|N^+>= 
g\bar{u}(p')\gamma_5 u(f),
\end{equation}
where $f$ is the 4-momentum of the intermediate nucleon in the 
diagram of Fig. 2(a). For the 
clarity, we will take only that part of the nucleon
electromagnetic current which is related to  
the $F_1(Q^2)$ Pauli form factor:
\begin{eqnarray}
&&<N^+,f|J_{\mu}|N^+,p>=F_1^p(Q^2)\bar{u}(f)
\gamma_{\mu}u(p).
\end{eqnarray}

Now let us write the matrix 
elements for the contributions
of the Fig. 1, 2(a) diagrams to 
$\gamma^* p\rightarrow \pi^0 p$:
\begin{eqnarray}
Fig.~2a:&&M=gF_1^p(Q^2)\bar{u}(p')\gamma_5\frac{m+\hat{f}}{m^2-f^2}
\hat{\epsilon} u(p),\\
Fig.~1:&&M=\pm C_Ieg^*G_1(Q^2)\bar{u}(p')\gamma_5\frac
{\pm M+\hat{f}}{M^2-f^2}\nonumber\\
&&[k^2(\gamma\epsilon)
-(\gamma k)(k\epsilon)]u(p),
~ J^P=\frac{1}{2}^{\pm},\\
&&f=p+k=p'+q,~Q^2=-k^2,
\end{eqnarray}
where $\hat{a}\equiv (a\gamma)$, and again for the simplicity, 
we have taken only the part of the $\gamma^* N\rightarrow N^*$
vertex related to  
the $G_1(Q^2)$ form factor.

The general form of 
the $\gamma^* N\rightarrow N\pi$ matrix
element according to the definition of Ref. \cite{Devenish1} is:
\begin{eqnarray}
M=&&\bar{u}(p')\gamma_5 J\epsilon u(p),\\
J\epsilon= &&\frac{B_1(Q^2)}{2}
[(\gamma \epsilon)(\gamma k)-(\gamma k)(\gamma \epsilon)]+\nonumber\\
&&+2B_2(Q^2)\left[({\cal P}\epsilon)-({\cal P} k)\frac{k 
\epsilon}{k^2}\right]+\nonumber\\
&&+2B_3(Q^2)\left[(q\epsilon)-(q k)\frac{k 
\epsilon}{k^2}\right]-\nonumber\\
&&-B_5(Q^2)\left[(\gamma\epsilon)-(\gamma k)\frac{k 
\epsilon}{k^2}\right]+\nonumber\\
&&+B_6(Q^2)(\gamma k)\left[({\cal P}\epsilon)-({\cal P} k)\frac{k 
\epsilon}{k^2}\right]+\nonumber\\
&&+B_8(Q^2)(\gamma k)\left[(q\epsilon)-(q k)\frac{k 
\epsilon}{k^2}\right],\nonumber 
\end{eqnarray}
where ${\cal P}\equiv \frac{1}{2}(p+p')$.

Multipole amplitudes $E_{0+},~S_{0+},~M_{1-},~S_{1-}$ 
are related to the invariant amplitudes $B_i(Q^2)$
by \cite{Azn0}:
\begin{eqnarray}
&&E_{0+}=\tilde{E}_{0+}
\frac{\sqrt{(E+m)(E'+m)}}{8\pi W},\\
&&M_{1-}=\tilde{M}_{1-}
\frac{\sqrt{(E-m)(E'-m)}}{8\pi W},\\
&&S_{0+}=\tilde{S}_{0+}\frac{\sqrt{(E-m)(E'+m)}}{8\pi WQ^2},\\
&&S_{1-}=\tilde{S}_{1-}\frac{\sqrt{(E+m)(E'-m)}}{8\pi WQ^2},
\end{eqnarray}
\begin{eqnarray}
&&\tilde{E}_{0+}=(W-m)B_1-B_5,\nonumber\\
&&\tilde{M}_{1-}=-(W+m)B_1-B_5,\nonumber\\
&&\tilde{S}_{0+}=X(E+m)-\beta Y,\nonumber\\
&&\tilde{S}_{1-}=X'(E-m)+\beta Y',\nonumber
\end{eqnarray}
\begin{eqnarray}
&&\beta=\frac{W-E}{2}(t-m_{\pi}^2+Q^2)-Q^2(W-E'),\nonumber\\
&&X=Q^2B_1+(W-m)B_5+\nonumber\\
&&~~~~~~~~2W(E-m)\left(B_2-\frac{W+m}{2}B_6\right),\nonumber\\
&&Y=2B_3-B_2+(W+m)
\left(\frac{B_6}{2}-B_8\right),\nonumber\\
&&X'=-Q^2B_1+(W+m)B_5-\nonumber\\
&&~~~~~~~~2W(E+m)\left(B_2+\frac{W-m}{2}B_6\right),\nonumber\\
&&Y'=2B_3-B_2-(W-m)
\left(\frac{B_6}{2}-B_8\right).\nonumber
\end{eqnarray}
Now from Eqs. (59,62) it is easy to find
the contribution of the Fig. 2(a) diagram to 
$\gamma^* p\rightarrow \pi^0 p$:
\begin{eqnarray}
&&B_1(Q^2)=\frac{gF_1^p(Q^2)}{s-m^2},\\
&&B_2(Q^2)=2B_3(Q^2)=-B_1(Q^2),\\ 
&&B_5(Q^2)=B_6(Q^2)=B_8(Q^2)=0.
\end{eqnarray}
These relations coincide with the commonly used expressions
for the Born term which corresponds to the $s$-channel
nucleon exchange.

For the contribution of the Fig. 1 diagram to 
$\gamma^* p\rightarrow \pi^0 p$
we have:

\begin{eqnarray}
&&B_1(Q^2)=A,\\
&&B_2(Q^2)=2B_3(Q^2)=-B_1(Q^2),\\ 
&&B_6(Q^2)=B_8(Q^2)=0,\\
&&B_5(Q^2)=(-m\pm M)A,
~~ J^P=\frac{1}{2}^{\pm},\\
&&A\equiv C_I Q^2\frac{eg^*G_1(Q^2)}{M^2-s-iM\Gamma}.
\end{eqnarray}

From the relations (2-9,63-66,70-74), we 
find for
the $J^P={\frac{1}{2}}^+$ resonances:
\begin{eqnarray}
&&A_{\frac{1}{2}}=
2Q^2G_1(Q^2)b',\\
&&S_{\frac{1}{2}}=2(M+m)G_1(Q^2)b'\frac{|{\bf{k}}|}{\sqrt{2}},\\
&&b'=-e\frac{C_Ig^*}{a}\frac{\sqrt{(E-m)(E'-m)}}{8\pi M\Gamma}.\nonumber
\end{eqnarray}
For the $J^P={\frac{1}{2}}^-$ resonances:
\begin{eqnarray}
&&A_{\frac{1}{2}}=
2Q^2G_1(Q^2)b',\\
&&S_{\frac{1}{2}}=-2(M-m)G_1(Q^2)b'\frac{|{\bf{k}}|}{\sqrt{2}},\\
&&b'=-e\frac{C_Ig^*}{a}\frac{\sqrt{(E+m)(E'+m)}}{8\pi M\Gamma}.\nonumber
\end{eqnarray}

Now from the comparison with Eqs. (1,46-49) and using
Egs. (54,56),  we find:

\begin{equation}
\zeta=-sign(g^*/g).
\end{equation}

Using the results of Ref. \cite{Devenish1},
it can be shown that the same relation between $\zeta$
and $sign(g^*/g)$ is correct also for the resonances
with $J\geq \frac{3}{2}$, if the $\pi NN^*$ vertices
are defined in the following way:
for the resonances with $J^P=\frac{3}{2}^+,\frac{5}{2}^-$,...
\begin{eqnarray}
<N|J_{\pi}(0)|N^{*+}>=C_Ig^*\bar{u}(p')p'_{\nu_1}
...p'_{\nu_l}u^{\nu_1...\nu_l}(p^*),
\end{eqnarray}
and for the resonances with $J^P=\frac{3}{2}^-,\frac{5}{2}^+$,...
\begin{equation}
<N|J_{\pi}(0)|N^{*+}>=-C_Ig^*\bar{u}(p')p'_{\nu_1}
...p'_{\nu_l}\gamma_5u^{\nu_1...\nu_l}(p^*),
\end{equation}
where $l=J-\frac{1}{2}$ and $u^{\nu_1...\nu_l}(p^*)$
is the generalized Rarita-Schwinger spinor.

\section{$\gamma^*p\rightarrow N^*$ helicity
amplitudes in nonrelativistic quark model}

In this Section we will present explicit formulas
for the calculation of the $\gamma^*p\rightarrow N^*$ helicity
amplitudes in nonrelativistic quark model.
We will present also in explicit form wave functions
for the N, $P_{11}(1440)$, $S_{11}(1535)$, and final
formulas for the 
$\gamma^*p\rightarrow P_{11}(1440),S_{11}(1535)$ 
helicity amplitudes.

In nonrelativistic quark model,
the matrix elements which enter Eqs. (30-32)
can be written in the form:
\begin{eqnarray}
&<N^{*+},S^*_z|J_{t,l}|N^+,S_z>=\\
&{\sum_i}\int d{\bf{q}}_{\lambda}d{\bf{q}}_{\rho}<S'^i_{z}
|J^i_{t,l}|S^{i}_{z}>\Phi_N({\bf{q}}_{\lambda},{\bf{q}}_{\rho})
\Phi_{N^*}({\bf{q}}'_{\lambda},{\bf{q}}'_{\rho})\nonumber\\ 
&\equiv3\int d{\bf{q}}_{\lambda}d{\bf{q}}_{\rho}<S'^{3}_{z}
|J^3_{t,l}|S^{3}_{z}>\Phi_N({\bf{q}}_{\lambda},{\bf{q}}_{\rho})
\Phi_{N^*}({\bf{q}}'_{\lambda},{\bf{q}}'_{\rho}),\nonumber 
\end{eqnarray}
where $\Phi_{N}({\bf{q}}_{\lambda},{\bf{q}}_{\rho})$,
$\Phi_{N^*}({\bf{q}}'_{\lambda},{\bf{q}}'_{\rho})$
are the radial parts of the $N,N^*$ wave functions, and
according to Eqs. (30-32,35,36), we have made notations:
$J_t\equiv\frac{{\bf{J}}_x+i{\bf{J}}_y}{\sqrt{2}}$, $J_l\equiv J_0$.
In the last part of Eq. (82), it is supposed that photon
interacts with the 3-rd quark.
In the calculations, it is convenient to define
the 3-rd quark momentum in the $N$ and $N^*$ in the form:
\begin{eqnarray}
&{\bf{q}}_3={\tilde{\bf{Q}}}-\frac{{\bf{k}}}{2},\\
&{\bf{q}}'_3={\tilde{\bf{Q}}}+\frac{{\bf{k}}}{2},
\end{eqnarray}
where ${\bf{k}}$ is the photon 3-momentum directed
along the z-axis in the $N^*$ rest frame, and 
${\bf{q}}'_3={\bf{q}}_3+{\bf{k}}$.
The momenta, which have appropriate
symmetries under ${\bf{q}}_1,{\bf{q}}_2$ exchanges, are: 
\begin{eqnarray}
&{\bf{q}}_{\rho}=\frac{{\bf{q}}_1-{\bf{q}}_2}{\sqrt{2}},
~~~{\bf{q}}'_{\rho}={\bf{q}}_{\rho},\\
&{\bf{q}}_{\lambda}={\tilde{\bf{Q}}}_{\lambda}+\frac{{\bf{k}}}{\sqrt{6}},
~~~{\bf{q}}'_{\lambda}={\tilde{\bf{Q}}}_{\lambda}-\frac{{\bf{k}}}{\sqrt{6}},\\
&{\tilde{\bf{Q}}}_{\lambda}\equiv
\frac{{\bf{q}}_1+{\bf{q}}_2-2{\tilde{\bf{Q}}}}{\sqrt{6}}.
\end{eqnarray}

The matrix elements of the quark electromagnetic current have
the following form:
\begin{eqnarray}
&&<S^{'3}_{z}|J^3_{t}|S^{3}_{z}>=e_3\bar{u}(q'_3)
\frac{\vec{\gamma}_x+i\vec{\gamma}_y}{\sqrt{2}}u(q_3)\nonumber\\
&&=e_3\frac{\sqrt{2}}{m_q}
\phi_z^{'3}\left(\begin{array}{cc}
\tilde{Q}_x+i\tilde{Q}_y&
\frac{|{\bf{k}}|}{2}\\0&\tilde{Q}_x+i\tilde{Q}_y
\end{array}\right)\phi^{3}_{z},
\end{eqnarray}
\begin{equation}
<S^{'3}_{z}|J^3_{l}|S^{3}_{z}>=e_3\bar{u}(q'_3)
\gamma_0u(q_3)=e_3\phi^{'3}_{z}\left(\begin{array}{cc}
1&0\\0&1\end{array}\right)\phi^{3}_{z},
\end{equation}
where $m_q$ is the quark mass, and $\phi_z,\phi'_z$
are the quark spin wave functions in the initial
and final states.

The radial part of the nucleon wave function with harmonic oscillator
potential is
\begin{eqnarray}
&\Phi_N({\bf{q}}_{\lambda},{\bf{q}}_{\rho})=
\frac{1}{\pi^{3/2}\beta^3}
exp{\left(-\frac{{\bf{q}}^2_{\rho}+{\bf{q}}^2_{\lambda}}{2\beta^2}\right)},
\end{eqnarray}
where $\beta$ is the
harmonic-oscillator parameter.

In the classification over group $SU(6)\times O(3)$, 
the proton is the member of the octet from
the multiplet $[56,0^+]$.
The spin part of the proton wave function is:

\begin{equation}
|p,S_z>=\frac{1}{\sqrt{2}}\left(|p>_{\rho}
|\frac{1}{2},S_z>_{\rho}+|p>_{\lambda}
|\frac{1}{2},S_z>_{\lambda}\right),
\end{equation}
where $|p>_{\rho,\lambda}$ are 
$\rho$ and $\lambda$ -type flavor  wave functions for the octet:
\begin{eqnarray}
&&|p>_{\rho}=
\frac{1}{\sqrt{2}}(udu-duu),\\
&&|p>_{\lambda}=
\frac{1}{\sqrt{6}}(2uud-udu-duu),
\end{eqnarray}
and $|\frac{1}{2},S_z>_{\rho,\lambda}$ are
$\rho$ and $\lambda$ -type spin $\frac{1}{2}$ wave functions:
\begin{eqnarray}
&&|\frac{1}{2}\frac{1}{2}>_{\rho}=
\frac{1}{\sqrt{2}}(\uparrow\downarrow\uparrow
-\downarrow\uparrow\uparrow),\\
&&|\frac{1}{2}\frac{1}{2}>_{\lambda}=
\frac{1}{\sqrt{6}}(2\uparrow\uparrow\downarrow-
\uparrow\downarrow\uparrow
-\downarrow\uparrow\uparrow),\\
&&|\frac{1}{2}-\frac{1}{2}>_{\rho}=
\frac{1}{\sqrt{2}}(\uparrow\downarrow\downarrow
-\downarrow\uparrow\downarrow),\\
&&|\frac{1}{2}-\frac{1}{2}>_{\lambda}=
-\frac{1}{\sqrt{6}}(2\downarrow\downarrow\uparrow-
\downarrow\uparrow\downarrow
-\uparrow\downarrow\downarrow).
\end{eqnarray}

In the explicit form this gives:
\begin{eqnarray}
|p,\frac{1}{2}>=&&\frac{1}{\sqrt{18}}
(2u\uparrow u\uparrow d\downarrow-
u\uparrow u\downarrow d\uparrow-u\downarrow u\uparrow 
d\uparrow\nonumber\\
&&-u\uparrow d\uparrow u\downarrow+2
u\uparrow d\downarrow u\uparrow-u\downarrow d\uparrow 
u\uparrow\\
&&-d\uparrow u\uparrow u\downarrow-
d\uparrow u\downarrow u\uparrow+2d\downarrow u\uparrow 
u\uparrow),\nonumber
\end{eqnarray}
\begin{eqnarray}
|p,-\frac{1}{2}>=&&-\frac{1}{\sqrt{18}}
(2u\downarrow u\downarrow d\uparrow-
u\downarrow u\uparrow d\downarrow-u\uparrow u\downarrow 
d\downarrow\nonumber\\
&&-u\downarrow d\downarrow u\uparrow+2
u\downarrow d\uparrow u\downarrow-u\uparrow d\downarrow 
u\downarrow\\
&&-d\downarrow u\downarrow u\uparrow-
d\downarrow u\uparrow u\downarrow+2d\uparrow u\downarrow 
u\downarrow).\nonumber
\end{eqnarray}

\subsection{$\gamma^*p\rightarrow P_{11}(1440)$}
In the classification over group $SU(6)\times O(3)$, 
we will consider the 
resonance $P_{11}(1440)$ as the   
member of the octet from the multiplet $[56,0^+]_r$.
The spin part of the $P_{11}(1440)$ wave function
in this case is the same as for the nucleon (91,98,99), and
the radial part is:
\begin{eqnarray}
&&\Phi_{N^*}({\bf{q}}'_{\lambda},{\bf{q}}'_{\rho})=\\
&&\frac{1}{\sqrt{3}\pi^{3/2}\beta^3}
\left(3-\frac{{\bf{q}}'^2_{\rho}+{\bf{q}}'^2_{\lambda}}{\beta^2}\right)
exp{\left(-\frac{{\bf{q}}'^2_{\rho}+{\bf{q}}'^2_{\lambda}}{2\beta^2}\right)}
\nonumber.
\end{eqnarray}
The $\gamma^*p\rightarrow P_{11}(1440)$ helicity amplitudes
obtained using the definitions (30,32) and the results presented in
this Section are:
\begin{eqnarray}
&&{\cal{A}}_{\frac{1}{2}}=
-\sqrt{\frac{2\pi\alpha}{K}}\frac{|{\bf{k}}|^3}{6\sqrt{6}m_q\beta^2}
 e^{-\frac{{\bf{k}}^2}{6\beta^2}},\\
&&{\cal{S}}_{\frac{1}{2}}=
-\sqrt{\frac{2\pi\alpha}{K}}\frac{{\bf{k}}^2}{6\sqrt{3}\beta^2}
e^{-\frac{{\bf{k}}^2}{6\beta^2}}.
\end{eqnarray}

\subsection{$\gamma^*p\rightarrow S_{11}(1535)$}

In the classification over group $SU(6)\times O(3)$, 
we will consider the 
resonance $S_{11}(1535)$ as the ${}^2 8_{\frac{1}{2}}$  
member of the multiplet $[70,1^-]$.
The spin part of the $S_{11}(1535)$ wave function
in this case is:
\begin{eqnarray}
&&\mid N^{*+},\frac{1}{2}>=-
\frac{1}{2\sqrt{3}}\left(|p>_{\lambda}X_{\lambda}+
|p>_{\rho}X_{\rho}\right),\\
&&X_{\lambda}\equiv -|\frac{1}{2}\frac{1}{2}>_{\lambda}|\lambda,0>+
\sqrt{2}\mid\frac{1}{2}-\frac{1}{2}>_{\lambda}|\lambda,+1>\nonumber\\
&&+|\frac{1}{2}\frac{1}{2}>_{\rho}|\rho,0>-
\sqrt{2}|\frac{1}{2}-\frac{1}{2}>_{\rho}|\rho,+1>,\\
&&X_{\rho}\equiv|\frac{1}{2}\frac{1}{2}>_{\rho}|\lambda,0>-
\sqrt{2}|\frac{1}{2}-\frac{1}{2}>_{\rho}|\lambda,+1>\nonumber\\
&&+|\frac{1}{2}\frac{1}{2}>_{\lambda}|\rho,0>-
\sqrt{2}|\frac{1}{2}-\frac{1}{2}>_{\lambda}|\rho,+1>,
\end{eqnarray}
where
\begin{eqnarray}
&&|\lambda,0>\equiv {\bf{q}}'_{\lambda,z},~~~
|\lambda,+1>\equiv 
-\frac{{\bf{q}}'_{\lambda,x}+i{\bf{q}}'_{\lambda,y}}{\sqrt{2}},\\
&&|\rho,0>\equiv {\bf{q}}'_{\rho,z},~~~
|\rho,+1>\equiv 
-\frac{{\bf{q}}'_{\rho,x}+i{\bf{q}}'_{\rho,y}}{\sqrt{2}}.
\end{eqnarray}
The radial part of the $S_{11}(1535)$ wave function is:
\begin{equation}
\Phi_{N^*}({\bf{q}}'_{\lambda},{\bf{q}}'_{\rho})=
\frac{\sqrt{2}}{\pi^{3/2}\beta^4}
exp{\left(-\frac{{\bf{q}}'^2_{\rho}+{\bf{q}}'^2_{\lambda}}{2\beta^2}\right)}.
\end{equation}
The $\gamma^*p\rightarrow S_{11}(1535)$ helicity amplitudes
obtained according to the definitions (30,32)
and the results of this Section are:
\begin{eqnarray}
&{\cal A}_{1/2}=\sqrt{\frac{8\pi\alpha}{K}}
\frac{\beta}{6m_q}
\left(1+\frac{\bf{k}^2}{2\beta^2}\right)
exp\left(-\frac{\bf{k}^2}{6\beta^2}\right),\\
&{\cal S}_{1/2}=
\sqrt{\frac{4\pi\alpha}{K}}
\frac{|\bf{k}|}{6\beta}
exp\left(-\frac{\bf{k}^2}{6\beta^2}\right).
\end{eqnarray}

\section{Information on the signs of the $\pi N N^*$ coupling
constants available in quark model}

In the framework of quark model,
the signs of the $\pi N N^*$ coupling
constants were found in Ref. \cite{Azn70}
for the resonances of the multiplet $[70,1^-]$
using an approach based on PCAC. There are also results
on the sign of the $\pi N P_{11}(1440)$ coupling constant
obtained in Refs. \cite{Capstick,Azn56}, respectively
in ${}^3P_0$ model and using PCAC;
they coincide with each other. 
Quark model predictions for the
$\gamma^*p\rightarrow N^*$ helicity amplitudes
are presented in Refs.  \cite{Azn70,Capstick,Azn56}
taking into account 
the $\pi N N^*$ signs obtained in these papers.

However, traditionally, quark model predictions for the
$\gamma^*p\rightarrow N^*$ helicity amplitudes
are presented with the common sign fixed
by taking the sign of $A_{1/2}$ at $Q^2=0$ equal
to that of the amplitude extracted from 
experimental data. Sometimes, such definition
of the sign can bring to confusing and wrong results,
as we will demonstrate below on the example of
the $P_{11}(1440)$ resonance.

\begin{figure}[ht]
\includegraphics[width=9.5cm, bb=20 160 590 600]{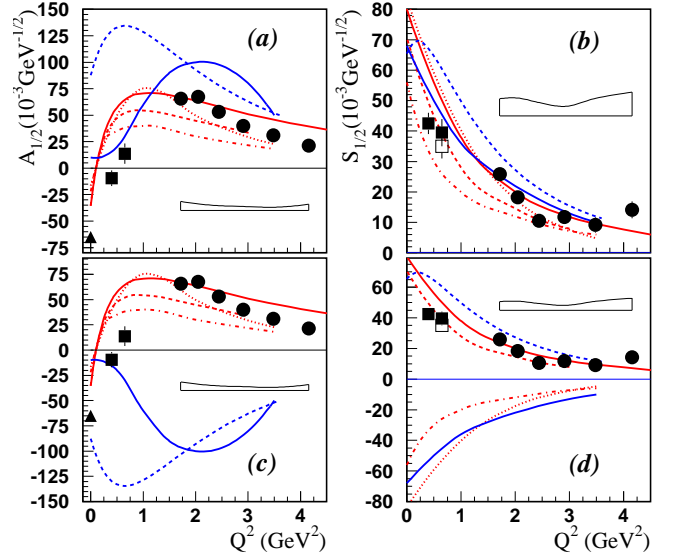}
\caption{Helicity amplitudes for the $\gamma^* p\rightarrow 
P_{11}(1440)$ transition. The full circles are the data
extracted from the JLab-CLAS 
$\vec{e}p\rightarrow ep\pi^0,en\pi^+$ data
\cite{Joo1,Joo2,Joo3,Egiyan,Park,Azn04,Azn065,AznRoper}. 
The bands present model uncertainties of the data.
The full triangle at $Q^2=0$ is the RPP estimate \cite{PDG}.
The red curves
correspond to the predictions of the light-front
relativistic quark models: 
dashed - Ref. \cite{Capstick},
solid  - Ref. \cite{Azn56},
dotted  - Ref. \cite{Weber},
dashed-dotted  - Ref. \cite{Simula}.
The blue  curves are the results
obtained via nonrelativistic calculations:
solid  - Ref. \cite{Warns},
dashed -  Ref. \cite{Santopinto}.
The plots (c,d) present the predictions in that form
as they are given
in the papers. In plots (a,b), all results 
are presented with correct signs.
}
\label{fig:fig3}
\end{figure}

In Fig. 3, we present the results for the 
$\gamma^*p\rightarrow P_{11}(1440)$ helicity amplitudes
extracted from the JLab-CLAS data on
the $\vec{e}p\rightarrow ep\pi^0,en\pi^+$ reactions
\cite{Joo1,Joo2,Joo3,Egiyan,Park,Azn04,Azn065,AznRoper}
in comparison with the quark model predictions obtained
in the light-front dynamics
\cite {Capstick,Azn56,Weber,Simula}
and via nonrelativistic calculations \cite{Warns,Santopinto}.
For clear understanding of the  results
presented in Fig. 3, it is important to  note that the 
$A_{1/2}$ amplitude
found with the same $N$ and $P_{11}(1440)$
wave functions via nonrelativistic calculations
and in the light-front relativistic approaches have at $Q^2=0$ opposite 
signs. For the first time this was mentioned in Ref. \cite 
{Capstick}.  However, in the presentation of the results
in traditional way (see plots (c,d)), 
the amplitudes $A_{1/2}$ at $Q^2=0$ from Refs.
\cite {Capstick,Azn56,Weber,Simula}
and  \cite{Warns,Santopinto} appear with the same sign.
As a result, for higher $Q^2$, we have strong disagreement
between  the corresponding predictions, and
the quark model predictions  \cite{Warns,Santopinto}  
strongly disagree with the data extracted from experiment.

In plots (a,b), we present all results with the $\pi N P_{11}(1440)$
sign found in Refs. \cite{Capstick, Azn56}.
We have also corrected the relative signs between the
$A_{1/2}$ and $S_{1/2}$ amplitudes from
Refs. \cite{Weber,Simula,Santopinto}.
It can be seen that for $Q^2>0.4~GeV^2$, this results in
good agreement with experiment for the signs of amplitudes
from all approaches \cite {Capstick,Azn56,Weber,Simula,
Warns,Santopinto}; we have also better
agreement of the results obtained in different approaches
with each other. 

For the resonances from the multiplet $[70,1^-]$:
N(1520)D$_{13}$, N(1535)S$_{11}$, 
$\Delta(1620)$S$_{31}$,
N(1650)S$_{11}$,
N(1700)D$_{13}$, and $\Delta(1700)$D$_{33}$,
the signs found in traditional way and in Ref. \cite{Azn70}
coincide with each other for all resonances,
except N(1700)D$_{13}$. With the $\pi N N^*$ 
coupling constants found in
Ref.  \cite{Azn70},
nonrelativistic 
quark model prediction for the 
$A_{1/2}(\gamma^*p\rightarrow N^*)$ amplitude
at $Q^2=0$ is positive for
the resonances N(1535)S$_{11}$,
$\Delta(1620)$S$_{31}$,
N(1650)S$_{11}$,
N(1700)D$_{13}$, and $\Delta(1700)$D$_{33}$,
and negative for N(1520)D$_{13}$.

\section{Summary}

By performing covariant calculations of the 
resonance (Fig. 1) and Born terms (Fig. 2) contributions
to $\gamma^* p\rightarrow \pi N$, we have found
the relations between the following definitions of
the $\gamma^* p\rightarrow N^{*+}$ helicity amplitudes
$A_{1/2},A_{3/2},S_{1/2}$:

(i) the definition through the $\gamma^* p\rightarrow \pi N$ multipole
amplitudes which is commonly used for the extraction of
$A_{1/2},A_{3/2},S_{1/2}$ from experimental data on 
$\gamma^* p\rightarrow \pi N$;

(ii) the definitions through the hadron electromagnetic current
and  the $\gamma^* p\rightarrow N^*$ form factors
which are used in theoretical calculations.

These relations include the relative sign between 
the coupling constants $g^*$ and $g$ which we have defined 
explicitly for the vertices 
$\pi NN^*$ and $\pi NN$. For the proper comparison
of theoretical predictions with the results
extracted from experimental data,
it is important to have
both quantities, the $\gamma^* p\rightarrow N^{*+}$ amplitudes
and the sign of the ratio $(g^*/g)$,  calculated within 
the same theoretical approach. 
This is demonstrated on the example of the 
$\gamma^*p\rightarrow P_{11}(1440)$ transition.

The performed calculations allowed us also to check and present
the formulas for the $\gamma^* p\rightarrow N^{*+}$ helicity amplitudes
through the hadron electromagnetic current
and  the $\gamma^* p\rightarrow N^*$ form factors
which give the correct relative sign between the longitudinal $S_{1/2}$
and transverse $A_{1/2},A_{3/2}$ amplitudes,
consistent with that for  the amplitudes extracted from experiment. 
To avoid sources of mistakes 
in the calculation of this sign, we give 
definitions of all quantities which enter the formulas.  

For completeness,
we have presented formulas for 
the calculations of the $\gamma^*N\rightarrow N^*$ helicity
amplitudes in nonrelativistic quark model along with
the N, $P_{11}(1440)$, $S_{11}(1535)$
wave functions and final
formulas for the
$\gamma^*p\rightarrow P_{11}(1440),S_{11}(1535)$
helicity amplitudes.

Taking into account the available information
on the signs  of the ratios $(g^*/g)$ in the framework
of quark model,
we have presented, in the case of 
nonrelativistic quark model, the signs of the $A_{1/2}$ amplitudes
for the resonances from the multiplet $[70,1^-]$.
This may be useful for the correct presentations of
the signs of
the helicity amplitudes $A_{1/2},A_{3/2},S_{1/2}$  in quark model.

\section{Acknowledgements}

I. G. Aznauryan is grateful to Ya. Azimov and 
J. L. Goity for useful discussions.

This work was supported in part by the U.S. Department
of Energy and the National Science Foundation.
Jefferson Science Associates, LLC, operates Jefferson Lab
under U.S. DOE contract DE-AC05-060R23177.

\end{document}